# Direct measurements of laser light aberration from the ARTEMIS geostationary satellite through thin clouds

Volodymyr Kuzkov [1*], Sergii Kuzkov [1], Zoran Sodnik [2]
1. Main Astronomical Observatory of National Academy of Sciences, Kyiv, Ukraine,
2. European Space Research and Technology Centre, ESA, Noordwijk, The Netherlands
* *Email* address: kuzkov@mao.kiev.ua

A precise ground based telescope system was developed for laser communication experiments with the geostationary satellite ARTEMIS of ESA. Precise tracking of the satellite was realized by using time resolved coordinates of the satellite. During the experiments, the time propagation of laser signal from the satellite and the point-ahead angle for the laser beam were calculated. Some laser experiments though thin clouds were performed. A splitting of some images of the laser beam from the satellite along declination and right ascension coordinates of telescope could be observed through thin clouds. The splitting along the declination coordinate may be interpreted as refraction in the atmosphere. The splitting along the right ascension coordinate is equivalent to the calculated point-ahead angle for the satellite. We find out that a small part of laser beam was observed ahead of the velocity vector in the point where the satellite would be after the laser light from the satellite reaches the telescope. These results are in accordance with the theory of relativity for aberration of light during transition from immovable to movable coordinate systems. We directly observed laser light aberration as result of moving of satellite with angular velocity close to Earth rotation.

## 1. INTRODUCTION

In July 2001, ESA's Geostationary Earth Orbiting (GEO) Advance data-Relay and Technology Mission Satellite (ARTEMIS) was launched with a laser communication terminal OPALE on-board. Laser communication sessions were successfully performed between ARTEMIS and the low Earth orbiting SPOT-4 satellite. The results are presented in Talker-Nielsen T. and Oppenhauser G., (2002). Multiple laser communication experiments were also performed between ARTEMIS and ESA's optical ground station (OGS) with the results presented in Reyes M. et al., (2002); Romba J. et al., (2004) and Reyes M. et al., (2005). ESA's OGS 1-m telescope is located at Teide observatory (Canary Islands) at altitude of 2400-m. Successful laser communication experiments between Low Earth Orbiting (LEO) Kirari satellite and NICT optical ground station (KODEN) were realized with results presented in Toyoshima M. et al., (2012).

Laser communication systems have advantages in comparison with radio frequency communication systems. There is strong interest in using of laser systems for communications between ground stations and space missions through the atmosphere.

The Main Astronomical Observatory (MAO) of the National Academy of Science of Ukraine developed a 0.7-m telescope based system (MAO OGS) at the altitude of 190-m for laser communication experiments with the ARTEMIS satellite. The tracking of the satellite was realized by using time resolved coordinates of the satellite. A compact device called LACES (Laser atmosphere and Communication Experiments with Satellites) equipped with a laser module, pointing and tracking cameras, and other optical and mechanical components, was developed to be installed into the Cassegrain focus of the telescope with the description presented in Kuzkov V. et al., (2012). Multiple short-time laser communication experiments between MAO OGS and ARTEMIS were performed with results presented in Kuzkov S. et al., (2013). Some laser experiments we performed in cloudy conditions. The tentative results are presented in Kuzkov V. et al., (2014).

The stellar aberration, as the result of motion of the Earth in its orbit around the Sun, was discovered by James Bradley with results presented in Bradley J., (1727).

In our paper we present the results of direct laser light aberration measurements calculated during the movement of the satellite with an angular velocity close to angular rotation of the Earth. We do not know any other examples of direct observations of light aberration of space objects as the result of moving with velocity close to angular rotation of the Earth.

In Section 2 we describe the conditions of observations, while in Section 3 we present details of our observations and calculations. In Section 4 we explain the results and provide the analysis. Finally, in Section 5 we present a brief discussion and summary of our results. In Supplement we present details of estimation of possible influence of the atmosphere turbulence on the observed results.

## 2. CONDITIONS OF OBSERVATIONS

By the time of observations, the inclination of orbital plane of ARTEMIS towards equator plane of the Earth was about 10.341 degrees. The satellite's inclination increases in time because the satellite is not controlled in the North-South direction. As a result, additional deviations of satellite coordinates occur along declination and right ascension (hour angle) positions of the telescope.

The calculations of the satellite orbit coordinates were performed by two programs. The first program uses geocentric (X, Y, Z, Vx, Vy, Vz ) data for the fixed moment of time. The second program uses a two-line obit data obtained from the NORAD cataloger. On 26 October 2011, three laser sessions with ARTEMIS (19h, 20h, and 21h UTC) were recorded in partly clouded conditions.

The LACES pointing digital camera has a CMOS sensor of 3072×2048 pixels. It works with a focal reducer and its effective focal length is close to 5.425 m. The tracking CCD camera has a 752×582 pixel censor with 16 bit digital signal from each pixel and works at the telescope focus close to 10.85 m. Pixel sizes of the tracking CCD camera along X (right ascension or $\alpha$) and Y (declination or $\delta$) coordinate axes are 8.6 µm × 8.3 µm respectively. The tracking CCD camera was used in the observations with a 2×2 pixel binning. The calculated pixel scale in the focal plane was 0.327 arc-sec per pixel for X($\alpha$), and 0.316 arc-sec per pixel for Y($\delta$). The field of view of the tracking CCD camera is 2.05×1.53 arc min. The analysis of obtained images was performed with the Maxim DL-5Pro program. The images of a star moving near the ARTEMIS position for the current night of observations are presented in Fig. 1.

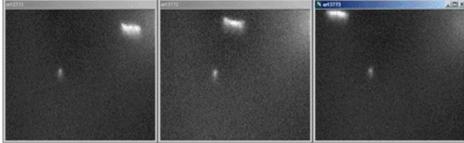

Fig. 1. The star moving near ARTEMIS position. Exposure was 1s.

The OPALE terminal beacon transmits a laser beam with a beam divergence of 155 arc-sec and wavelength band $\lambda = 797 - 808$ nm. OPALE performs a spiral scanning in the direction of the MAO OGS position with a maximum radius of 3.7 m-rad from the initial pointing direction. The duration of illumination of MAO OGS is near 1s. During every session two scans are performed with a total duration of 3+3 minutes. OPALE also has a narrow (7.5 µrad) laser communication beam at $\lambda = 819$ nm. During the current set of sessions this beam was not activated.

## 3. METHODOLOGY and EXPERIMENTAL MEASUREMETS

### 3.1 Session 1 (19 UTC )

Cloudy conditions occurred shortly before the start of the session. We performed automatic tracking of the satellite by our CCD camera. Exposure time was 2 s. We recorded images of the laser beam through clouds (Fig. 2).

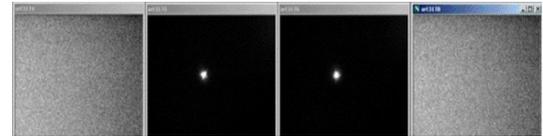

Fig. 2. Images art 3174 – art 3178. Laser beacon from OPALE through clouds: art 3175 – art 3176

A splitting of the laser beam could be observed for the image "art 3175" (Fig. 3). A splitting of laser beam on three components (A, B and C) was observed for this image. The angular splitting between A and C components was:
$\Delta X(\alpha)$ = 6 pixels or 1.962 arc-sec in the right ascension (hour angle) direction. The angular splitting along declination was $\Delta Y(\beta)$ = 3 pixels or 0.948 arc-sec The calculated satellite orbit data for this moment of time is shown in the Table 1.

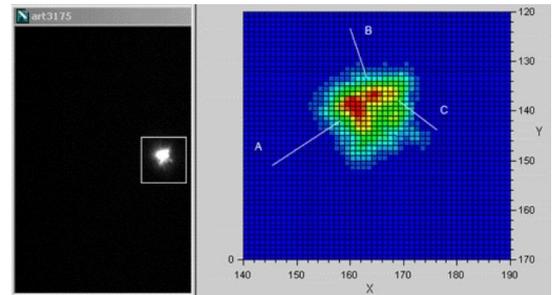

Fig. 3. The 2-D view of the image art3175

Table 1. The calculated satellite orbit data

| Date | UTC time | R.A. | Dec. | H.A. | Alt. | M. | Vel. | |
|---|---|---|---|---|---|---|---|---|
| | hh:mm:ss | hh:mm:ss | ° ' " | hh:mm:ss | ° | ° | H " / s | D |
| 26 Oct 2011 | 19:03:00 | 22:45:11.4 | -17 05 11 | 00:39:01.9 | 22 | 123 | -0.197 | 0.667 |
| 26 Oct 2011 | 19:04:00 | 22:46:12.3 | -17 04 31 | 00:39:01.1 | 22 | 123 | -0.196 | 0.678 |

Where: R.A. is right ascension; Dec. is declination; H.A. is hour angle; Alt. is altitude above the horizon; M. is an angle between the satellite, the MAO OGS and the Moon; H, D are angular velocities along hour angle and declination directions respectively.

Calculated satellite orbit data obtained from NORAD elements are in Table 2:

Table 2. Calculated satellite orbit data from NORAD data.

| Date | UTC time | R.A. | Dec. | L |
|---|---|---|---|---|
| mm/dd/yy | hh:mm:ss | hh:mm | ° ' | km |
| 10/26/11 | 19:03:00 | 22:45 | -17.05 | 39351 |

Where L is a distance from MAO OGS to the satellite.

By the end of the beacon scan, at the time period of 19h:07m:42s – 19h:08m:02s, the sky became clear and the satellite became visible in reflected sun light. Then the sky became overcast again. Comparative investigations of atmospheric instability and correlation between the motions of the images of close stars for ESA OGS and MAO telescope were performed some years ago with results presented in Kuzkov V. et al., (2005, 2008). By same methodology the coordinate measurements of positions of the satellite in reflected sunlight were performed to estimate the possible influence of turbulence on the results of observations (Table 5 in Supplement). A small drift can be seen (Fig. 13 and Fig. 14 in Supplement) which is negligible for the images obtained with 2 s exposures. If we remove the drift of the photometric centroids, the "jumps" would not exceed 0.4 arc-sec for the X($\alpha$) direction and 0.7 arc-sec for the Y($\delta$) direction. Interesting results of influence of atmosphere on laser communication are also presented in Giggenbach D., (2011) and Toyoshima M. et al., (2011).

## 3.2. Session 2 ( 20 UTC )

Cloudy conditions occurred before the start of the second session again. We performed automatic tracking of the satellite and registered images from the satellite beacon through clouds by our CCD camera. The exposure was 2 s. Coordinate data of the satellite for these images are below:

26 Oct  20:03:00   23 45 59.3  -16 05 37  00 38 23.8   23  134   - 0.113 1.308
26 Oct  20:04:00   23 46 60.0  -16 04 18  00 38 23.3   23  134   - 0.111 1.318
L-distance = 39256 km
26 Oct  20:06:00   23 49 01.2  -16 01 39  00 38 22.4   23  135  - 0.108 1.338
26 Oct  20:07:00   23 50 01.8  -16 00 18  00 38 22.0   23  135  - 0.106 1.348
L-distance = 39250 km.

The laser beacon of ARTEMIS through clouds can be seen in "art3307" and "art3308" images.

For the image "art3307" (Fig. 4), the angular splitting between components was:
$\Delta X(\alpha)$ = 0.948 arc-sec, $\Delta Y(\delta)$ = 3.270 arc-sec for A and B components;
$\Delta X(\alpha)$ = 1.962 arc-sec, $\Delta Y(\delta)$ = 0.948 arc-sec for B and C components;
$\Delta X(\alpha)$ = 0.981 arc-sec, $\Delta Y(\delta)$ = 4.424 arc-sec for A and C components.

For the image "art3308" (Fig. 5), the angular splitting between components was:
$\Delta X(\alpha)$ = 0.654 arc-sec, $\Delta Y(\delta)$ = 4.424 arc-sec for A and B components;
$\Delta X(\alpha)$ = 1.962 arc-sec, $\Delta Y(\delta)$ = 5.056 arc-sec for A and C components;
$\Delta X(\alpha)$ = 0.654 arc-sec, $\Delta Y(\delta)$ = 6.004 arc-sec for A and D components.

Sliced line of A and C components of the image art3308 can be seen in Fig. 6. Maximum of signal for A component is 47579 levels and for C component is 19486 levels. Background of clouds is 588 levels. Noise of the signals was approximately 60 levels. So the signal/noise ratio is 793 for A component and 325 for C component.

The next beacon peaks observed through clouds can be seen in image "art3350" ( Fig. 7) and image "art3351" ( Fig. 8).

For the image "art3350" (Fig. 7), the angular splitting between A and B components was: $\Delta X(\alpha)$ = 1.962 arc-sec, $\Delta Y(\delta)$ = 2.844 arc-sec.

For the image "art3351" (Fig.8), the angular splitting between A and B components was:  $\Delta X$ =1.962  arc-sec,  $\Delta Y(\delta)$ = 1.580 arc-sec.

After the end of the Session 2, coordinate measurements of ARTEMIS in reflected sunlight were also performed. 11 images (from "art3377" at 20h:09m:41s up to "art3387" at 20h:10m:21s) were calculated in the same way as in Session 1. The results of the calculations are presented in Fig. 15,16 (see the Supplement). The drifts are negligible during 2 s exposure. The maximum "jumps" do not exceed 0.5 arc-sec along the X axis and 1 arc-sec along the Y axis**.** drifts are negligible during 2 s exposure. The maximum "jumps" do not exceed 0.5 arc-sec along the X axis and 1 arc-sec along the Y axis**.**

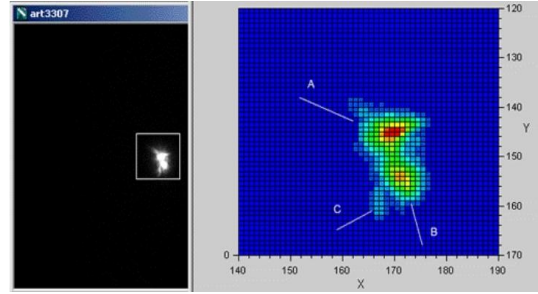
Fig. 4. The 2-D view of the image art3307

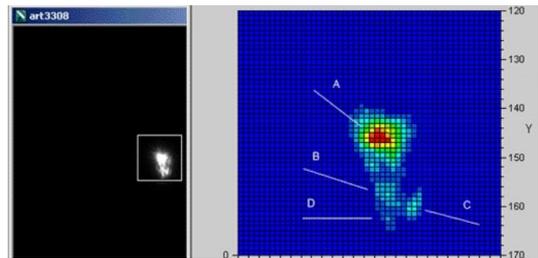
Fig. 5. The 2-D view of the image art3308

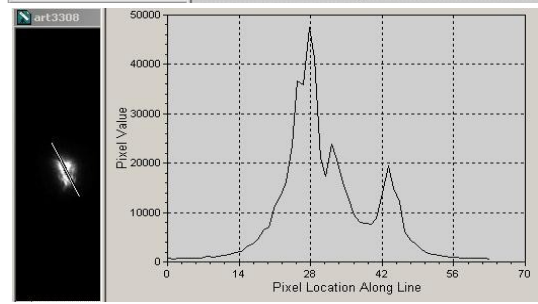
Fig. 6. Slice of A and C components of the image art3308

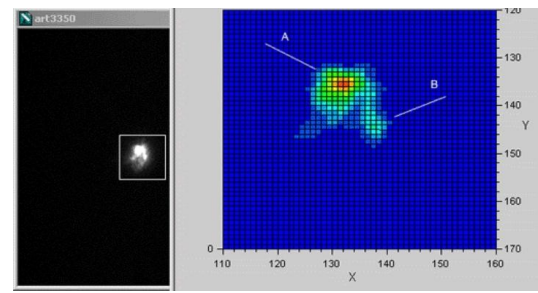
Fig. 7. The 2-D view of the image art3350

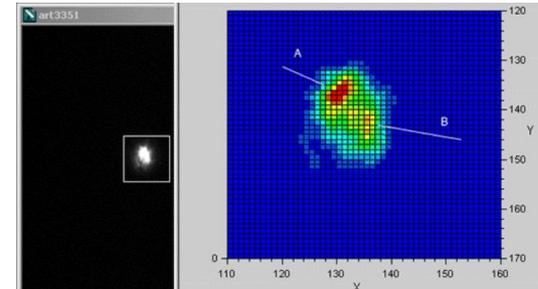
Fig. 8. The 2-D view of the image art3351

## 3.3. Session 3 ( 21 UTC )

Observations during the third session were performed in conditions of cirrus clouds. Exposure time was reduced to 1 s. The beacon maximum was observed in "art4331" and "art4388".
Calculated orbit data of the satellite are presented below:
26 Oct  21:00:00  00 43 22.4  -14 35 35  00 38 10.1  25 144  -0.004  1.835
26 Oct  21:01:00  00 44 22.6  -14 33 44  00 38 10.1  25 144  -0.002  1.844
L- distance = 39117 km.

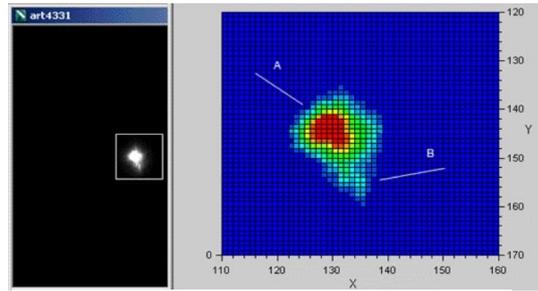

Fig. 9. The 2-D view of the image art 4331

Image "art4331" (Fig.9) was overexposed. The maximum splitting between components was: $\Delta X(\alpha)$ = 1.962 arc-sec and $\Delta Y(\delta)$ = 3.476 arc-sec. The image "art4388" was strongly overexposed. The area of overexposed pixels was approximately 9.5 arc-sec in X direction and 10.1 arc-sec in Y direction. A splitting of the laser beam was not observed due to the large area of overexposed pixels.

The coordinate measurements of ARTEMIS in reflected sunlight after the finish of Session 3 were also performed. 10 images from "art4466" at 21h:07m:22s up to "art4475" at 21h:07m:48s were calculated in the same way as in Session 1. The results of calculations are presented in Fig.17 and Fig. 18 (see the Supplement). The drifts are negligible during 1 s exposure. Maximum "jumps" of coordinates do not exceed 0.6 arc-sec along the X direction and 0.75 arc-sec along the Y direction.

## 4. RESULTS and ANALYSIS

As it can be seen from Figure 1, there is some angle between the X coordinate axis of the CCD censor and the right ascension (hour angle) direction. According to the images "art3771 – art3773", the calculated angle between the X coordinate axis of the sensor and the direction of star movement is 7.1°. Accordingly, the observed results $\Delta X(\alpha)$ (in arc-sec) must be multiplied by 1.0077 ($\cos^{-1}(7.1)$) to match the right ascension direction ($\Delta\alpha$). The images "art3175", "art 3307", "art3308", "art3350", "art3351" and "art4331" have approximately the same splitting along the X coordinate axis of the CCD censor with $\Delta X(\alpha)$ = 1.962 arc-sec It is equivalent to $\Delta\alpha$ = 1.977 arc-sec along the right ascension direction. The direction of the satellite movement is opposite to the direction of the star movement (Fig.1, Fig.10).

The time of propagation of laser light from the satellite to MAO OGS is determined by $T_{sig} = L \cdot V_{sig}^{-1}$ were $V_{sig}$ = C which is the velocity of light (299792 km·s$^{-1}$) in space. During the $T_{sig}$ time, the satellite is moving in space from the A point to the B point to the distance $L_{ab} = T_{sig} \cdot V$, were V is the velocity of a geostationary satellites in space. From NORAD elements of the satellite data we know that velocity of the satellite is 3.07 km·s$^{-1}$. The point-ahead angle Qf is determined as $Qf = L_{ab} \cdot L^{-1}$ or $Qf = V \cdot C^{-1}$. The result is Qf = 2.112 arc-sec. The point-ahead angle Qf can be calculated for orbit plane of the satellite. In our case, it is necessary to know the point-ahead angle at our telescope in the hour angle direction. We need the laser signal from our telescope to be sent directly into the point where satellite will be while the signal between our telescope and the satellite is passing. The orbit plane of the ARTEMIS has an inclination angle β =10° 20′ 28″ to Earth equator (right ascension) plane (Fig. 11) for the given night of observation. The projection of the velocity vector of the satellite has the same inclination. It is necessary to know the projections of Lab on the right ascension (hour angle) direction of the telescope for further correction of the hour angle (on the point-ahead angle) of the telescope in according with Tsig. The corrected point-ahead angles Qfαβ on hour angle direction are also determined and result is Qfαβ = 2.078 arc-sec. It is close to the point-ahead angle achieved by us. It is known that stars move with angular velocity $Q_f = 15 \cdot \cos(\delta)$ ″·s$^{-1}$. Polar stars have zero angular velocity. We calculated ahead angle (Qfαβ) for the right ascension direction. For different declinations (δ) of the satellite observed ahead angles will be Qfαβδ. The summarized results of observed and calculated ahead angles are presented in Table 3.

Table 3. Laser light aberration in X(α) direction

| # Art – image | Δα, arc-sec | L, km | Hvel, ″·s$^{-1}$ | Tsig, s | δ, deg | Qf, arc-sec | Qfαβδ, arc-sec |
|---|---|---|---|---|---|---|---|
| art3175 | 1.977 | 39351 | - 0.197 | 0.1313 | -17.086 | 2.112 | 1.986 |
| art3307 | 1.977 | 39256 | - 0.112 | 0.1309 | -16.081 | 2.112 | 1.997 |
| art3308 | 1.977 | 39256 | - 0.112 | 0.1309 | -16.081 | 2.112 | 1.997 |
| art3350 | 1.977 | 39250 | - 0.107 | 0.1309 | -16.016 | 2.112 | 1.997 |
| art3351 | 1.977 | 39250 | - 0.107 | 0.1309 | -16.015 | 2.112 | 1.997 |
| art4331 | 1.977 | 39117 | - 0.003 | 0.1305 | -14.577 | 2.112 | 2.011 |

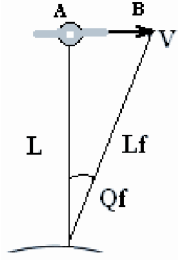 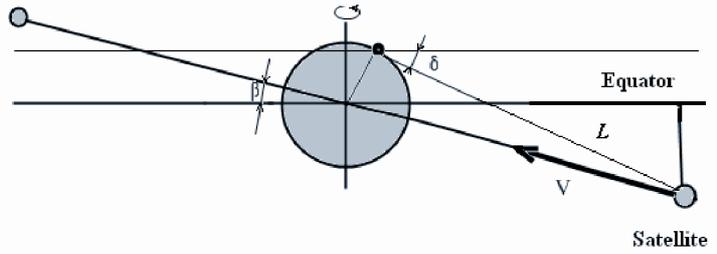

Fig.10. Ahead angle of the satellite   Fig. 11. Position of the satellite orbit to Equator

Observed values of Δα are close to the calculated point-ahead angles Qfαβδ obtained from the orbit coordinates. The accuracy of our measurements is estimated to be the size of two pixels during 2×2 binning and is 0.327 arc-sec per pixel along the X axis and 0.316 arc-sec per pixel along the Y axis. The calculated diffraction limit for our telescope is approximately 0.17 arc-sec. The duration of laser spot movement via the telescope is approximately 1 s. Velocity of light in the atmosphere in normal conditions is approximately 294000 km·s⁻¹. Time of propagation of light through 10 km for the near-earth layer of atmosphere is $3.4 \times 10^{-5}$ s. The influence of additional visible movement of satellite with velocity ± H arc-sec·s⁻¹ is negligible as the result of a small drift of the satellite along Tsig. The accuracy of tracking of the satellite for our telescope was measured for Session 1. Positions of laser beacon through clouds were measured for images "art3176" and "art3229" with times of recording at 19h:03m:00s and 19h:06m:33s respectively. Differences of positions between observed beacon in these images was ΔX(α) = 4.532 arc-sec and ΔY(δ) = 3.735 arc-sec. As the result, the drift is 0.021 arc-sec per second of time in the X(α) axis and 0.018 arc-sec per second of time in the Y(δ) axis. In our observations of propagation of laser light from the satellite to MAO OGS through clouds, we directly observed the positions of ARTEMIS that were equivalent to the calculated point-ahead angle for the satellite. In accordance with the theory of relativity, the aberration of light is changing the direction of light during the transition from immovable to movable coordinate systems. We have two coordinate systems. The first one (X', Y', and Z') is for the satellite (Fig. 12). The second coordinate system (X, Y, and Z) is for the telescope. The direction of the light θ' in the satellite coordinate system is determined by Eq. (1) in according with the description presented in C. Moller (1972).

$$\tan(\Theta') = \frac{\sin(\Theta) * \sqrt{1 - V^2/C^2}}{\cos(\Theta) + V/C} \qquad (1)$$

When the satellite tracking is performed, the X axis is parallel to X' and the Y axis is parallel to Y'. The satellite center is equivalent with the center of X', Y', Z' coordinates calculated for every time of observations. The center of our telescope CCD camera is equivalent to the center of X, Y, Z coordinates. As the result, the angle θ is 90° and the calculated angle θ' is determined by the equation (V<< C):

$$\tan(\Theta') = C/V \qquad (2)$$

In accordance with this Eq.(2), the calculated light aberration angle (the point-ahead angle) is determined as Δθ = θ − θ'. In our case, Δθ = 2.112 arc-sec We consider the velocity of a geostationary satellite equal to V=3.07 km·s⁻¹ while the velocity of light in space is 299792 km·s⁻¹. Thus, Δθ is equal to the point ahead angle Qf calculated previously.

The summarized results of splitting of observed images along Y(δ) direction are presented in Table 4. The possible interpretation is that this splitting is the result of Raman scattering of laser radiation (797–808 nm) on molecules of the atmosphere (perhaps water vapors in clouds) and large atmosphere refraction at altitudes of 22, 23 and 25 degrees above the horizon.

Table 4. Summarized results of splitting along Y(δ) direction

| # Art – image | Time h:m:s | ΔY(δ) arc-sec |
|---|---|---|
| art3175 | 19: 02: 56 | 0.948 |
| art3307 | 20: 03: 37 | 3.270 |
| art3307 | 20: 03: 37 | 4.424 |
| art3308 | 20: 03: 41 | 4.424 |
| art3308 | 20: 03: 41 | 5.056 |
| art3308 | 20: 03: 41 | 6.004 |
| art3350 | 20: 06: 30 | 2.844 |
| art3351 | 20: 06: 34 | 1.580 |
| art4331 | 21: 00: 35 | 3.476 |

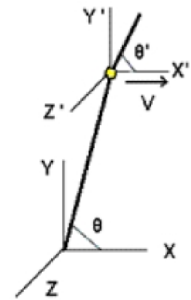

Fig.12. The aberration angle for the satellite and coordinate systems

## CONCLUSION

Laser experiments were performed with the ARTEMIS geostationary satellite in cloudy conditions. The images of propagation of radiation from the laser beacon of the satellite through thin clouds were recorded and analyzed. The splitting of laser beam was observed in some images. The positions of the components were calculated. The splitting of the laser beam was observed in declination and right ascension directions. The splitting along the declination direction may be interpreted as the result of Raman scattering on molecules and refraction in the atmosphere. The splitting in some observed images along the right ascension direction is equivalent to the calculated point-ahead angle for the satellite. We find out that a small part of laser beam was observed ahead of the velocity vector in the point where the satellite would be after the laser light from the satellite reaches our telescope. In accordance with our calculations signal/noise ratio for observed ahead point (Fig.6) is up to 325 value.

The calculated point-ahead angle $Q_f$ is equal to 2.112 arc-sec. The calculated light aberration angle $\Delta\theta$ is also equal to 2.112 arc-sec. These results are in accordance with the theory of relativity for aberration of light during the transition from immovable to the movable coordinate systems. During these experiments through thin clouds, we directly observed laser light aberration that is the result of movement of the satellite with a velocity close to the velocity of rotation of the Earth.

The achieved results open possibilities to the development of systems for on-line atmospheric turbulence compensation during ground-to-space laser communications through the atmosphere.

We hope that these observations and methodology of calculations will be useful in search of astronomical objects with directly observed light aberration.

## ACKNOWLEDGMENT

The authors would like to thank Vincenzo Caramia from Redu Space Services S.A., ESA, Redu, Belgium for his assistance during the preparation of laser experiments with ARTEMIS. We would like to thank S. P. Pukha and D.V. Volovyk for considerable contribution in development of our optical ground system. We would like also to thank Dr.P. F. Lazorenko for support in preparing of this paper.

## REFERENCES

Bradley, James, 1727. Account of a New Discovered Motion of the Fix's Stars. Phil. Trans. R. Soc.1727.35: 637–661.

Giggenbach D., 2011. Deriving an estimate for the Fried parameter in mobile optical transmission scenarios, Applied Optics, Vol. 50, No. 2,. Optical Society of America, Jan. 2011, pp. 222-226.

Kuzkov V., Andruk V., Sizonenko Yu., Sodnik Z., 2005. Investigation of Atmospheric Instability for Communication Experiments with ESA's Geostationary Satellite ARTEMIS. Kinematics and Physics of Celestial Bodies, Suppl., 2005, n 5, pp. 561-565.

Kuzkov V., Andruk V., Sodnik Z., Sizonenko Yu., Kuzkov S., 2008. Investigating the correlation between the motions of the images of close stars for laser communications experiments with the Artemis satellite, Kinematics and Physics of Celestial Bodies. 2008, vol. 24, Issue 1, pp. 56 – 62.

Kuzkov V., Volovyk D., Kuzkov S., Sodnik Z., Pukha S., 2010. Realization of laser experiments with ESA's geostationary satellite ARTEMIS. Space Science and Technology (ISSN 1561-8889), 2010, 16, N.2, pp. 65 – 69.

Kuzkov V., Volovyk D., Kuzkov S., Sodnik Z., Caramia V., Pukha S., 2012. Laser Ground System for Communication Experiments with ARTEMIS. Proceedings of International Conference on Space Optical Systems and Applications (ICSOS-2012), October 9-12, Corsica, France. 2012, 3.2., pp. 1– 9.

Kuzkov S., Sodnik Z., Kuzkov V., 2013. Laser communication experiments with ARTEMIS satellite. Proceedings of 64th International Astronautical Congress (IAC), 23-27 September 2013 in Beijing, China, IAC. 2013. B2.3.8., pp.1– 8.

Kuzkov V., Kuzkov S., Sodnik Z., Caramia V., 2014. Laser Experiments with ARTEMIS Satellite in Cloudy Conditions. Proceedings of International Conference on Space Optical Systems and Applications (ICSOS) 2014, Kobe, Japan, May 7 - 9. 2014. S 4-4., pp. 1– 8.

Moller C., 1972. The Theory of relativity. Second edition, Clarendon Press Oxford, 1972. § 2.11. The Doppler effect. Aberration of light.

Reyes M., Sodnik Z., Lopez P., Alonso A., Viera T., Oppenhauser G., 2002. Preliminary results of the in-orbit test of ARTEMIS with the Optical Ground Station, Proc. SPIE, 2002, vol. 4635, pp. 38 – 49.

Jose Romba, Zoran Sodnik, Marcos Reyes, Angel Alonso, Aneurin Bird, 2004. ESA's Bidirectional Space-to-Ground Laser Communication Experiments. Proc. SPIE, 2004 , vol. 5550, pp. 287– 298.

Reyes M., Alonso A., Chueca S., Fuensalida J., Sodnik Z., Cessa V., Bird A., 2005. Ground to space optical communication characterization. Proc. SPIE. 2005. N 5892, pp. 589202-1– 589202-16.

Tolker-Nielsen T., Oppenhauser G. 2002. In-orbit test result of an operational optical inter satellite link between ARTEMIS and SPOT4, SILEX. Proc. SPIE. 2002. 4635., pp. 1– 15.

Toyoshima M., Takenaka H., and Takayama Y., 2011. Atmospheric turbulence-induced fading channel model for space-to-ground laser communications links. Optics Express, 2011, 19 (17), 15965–15975.

Toyoshima M., Takenaka H., Shoji Y., Takayama Y., Koyama Y., and Kunimori H., 2012. Results of Kirari optical communication demonstration experiments with NICT optical ground station (KODEN) aiming for future classical and quantum communications in space. Acta Astronautica, 2012, 74, pp. 40 – 49.

**SUPPLEMENT**

During of all present observations the image recording time was 2 s. Before and after Sessions 1, 2 the exposure times were 2 s. During and after Session 3 the exposure time was only 1 s. Data of coordinate measurements of the satellite's positions after Session 1 in reflected sunlight are presented in Table 5 and Fig. 13–14. Small drift can be seen (direct lines in the Figures ), which is insignificant for our time of exposures. The results of measurements after Sessions 2–3 are presented in Fig.15–18 only for compact presentation.

Table 5. Session 1. Coordinate measurements of the satellite in Sun reflected light

| Image # | Time hh:mm:ss | Centroid X, pixels | Centroid Y, pixels | X - Xm, arc-sec | Y - Ym, arc-sec |
|---|---|---|---|---|---|
| art3246 | 19:07:42 | 190.391 | 133.412 | -0.828 | -0.427 |
| art3247 | 19:07:46 | 191.720 | 133.812 | -0.393 | -0.301 |
| art3248 | 19:07:50 | 193.330 | 132.483 | 0.133 | -0.721 |
| art3249 | 19:07:54 | 192.421 | 134.886 | -0.164 | 0.039 |
| art3250 | 19:07:58 | 193.814 | 136.491 | 0.291 | 0.546 |
| art3251 | 19:08:02 | 195.860 | 137.498 | 0.960 | 0.864 |
| Mean | | Xm: 192.923 | Ym: 134.764 | | |

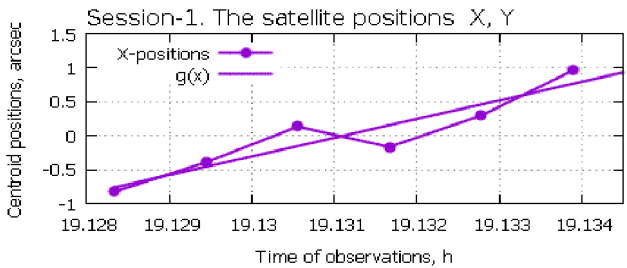

Fig. 13. Positions of the satellite along X direction

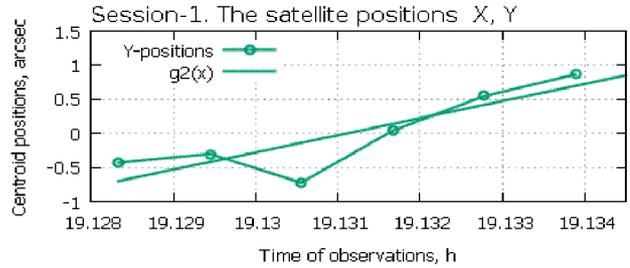

Fig. 14. Positions of the satellite along Y direction

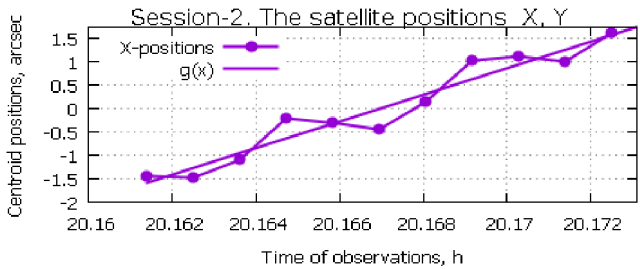

Fig. 15. Positions of the satellite along X direction

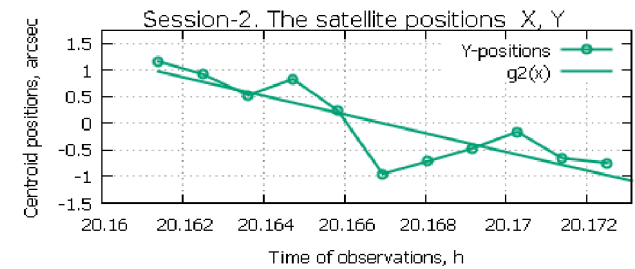

Fig. 16. Positions of the satellite along Y direction

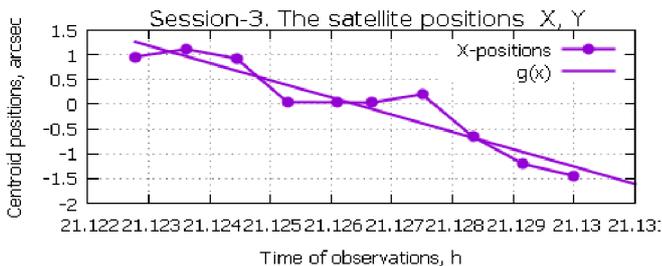

Fig. 17. Positions of the satellite along X direction

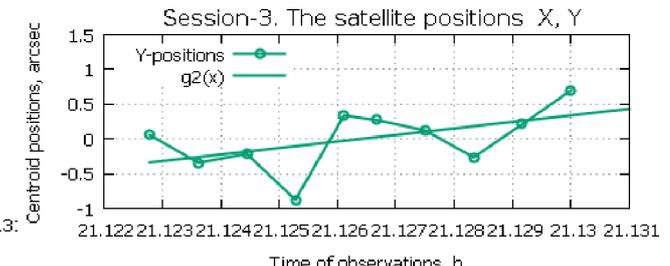

Fig. 18. Positions of the satellite along Y direction